\documentclass[10pt,conference]{IEEEtran}
\IEEEoverridecommandlockouts
\usepackage{amsmath,amsfonts}
\usepackage{algorithmic}
\usepackage{algorithm}
\usepackage{array}
\usepackage[caption=false,font=normalsize,labelfont=sf,textfont=sf]{subfig}
\usepackage{textcomp}
\usepackage{stfloats}
\usepackage[hyphens]{url}
\usepackage{hyperref}
\usepackage{float}
\usepackage{verbatim}
\usepackage{graphicx}
\usepackage{booktabs}
\usepackage{placeins}
\usepackage{cite}
\usepackage{amsmath,amssymb,amsfonts}
\usepackage{algorithmic}
\usepackage{graphicx}
\usepackage{textcomp}
\usepackage{color}
\usepackage{booktabs}
\usepackage[skins]{tcolorbox}
\usepackage{multirow}
\usepackage{listings}
\usepackage{xcolor}
\usepackage{colortbl}
\usepackage{caption}
\usepackage{subcaption}
\usepackage{pifont}
\usepackage{pgfplots}
\usepackage{threeparttable}
\usepackage[numbers,sort&compress]{natbib}
\usepackage{balance}
\usepackage{arydshln}
\usepackage{xspace}

\def\BibTeX{{\rm B\kern-.05em{\sc i\kern-.025em b}\kern-.08em
    T\kern-.1667em\lower.7ex\hbox{E}\kern-.125emX}}

\begin{document}

\newcommand{\todoc}[2]{{\textcolor{#1}{\textbf{[#2]}}}}
\newcommand{\todoblue}[1]{\todoc{blue}{\textbf{#1}}}
\newcommand{\todored}[1]{\todoc{red}{\textbf{#1}}}
\newcommand{\todopurple}[1]{\todoc{purple}{\textbf{#1}}}
\newcommand{\reply}[1]{\todoc{yellow}{\textbf{#1}}}
\newcommand{\modify}[1]{\textcolor{red}{#1}}

\definecolor{darkred}{rgb}{0.8, 0.1, 0.1}
\definecolor{darkgreen}{rgb}{0.1, 0.6, 0.1}
\newcommand{\better}[1]{\textcolor{darkred}{#1}}
\newcommand{\worse}[1]{\textcolor{darkgreen}{#1}}

\newcommand{\ytr}[1]{\todopurple{YTR: #1}}
\newcommand{\gu}[1]{\todoblue{gu: #1}}
\newcommand{\shen}[1]{\todoblue{shen: #1}}
\newcommand{\wan}[1]{{\color{cyan!70!blue}{[Wan: #1]}}}

\definecolor{lightgray}{gray}{0.9}

\title{Unraveling the Potential of Large Language Models in Code Translation: How Far Are We?}

\author{\IEEEauthorblockN{Qingxiao Tao$^*$, Tingrui Yu$^*$, Xiaodong Gu, Beijun~Shen$^{\dagger}$}\\
\IEEEauthorblockA{\textit{School of Software, Shanghai Jiao Tong University, Shanghai, China}\\
\{tao\_qingxiao, hzfsls, xiaodong.gu, bjshen\}@sjtu.edu.cn}}



\pagestyle{plain}
\maketitle

\begin{abstract}
While large language models (LLMs) exhibit state-of-the-art performance in various tasks, recent studies have revealed their struggle for code translation. This is because they haven't been extensively pre-trained with parallel multilingual code, which code translation heavily depends on. Moreover, existing benchmarks only cover a limited subset of common programming languages, and thus cannot reflect the full potential of LLMs in code translation. 
In this paper, we conduct a large-scale empirical study to exploit the capabilities and incapabilities of LLMs in code translation tasks. We first craft a novel benchmark called PolyHumanEval by extending HumanEval to a multilingual benchmark of 14 languages. With PolyHumanEval, we then perform over 110,000 translations with bleeding-edge code LLMs. The result shows LLMs' suboptimal performance on Python to other languages and the negligible impact of widely adopted LLM optimization techniques such as conventional pre-training and instruction tuning on code translation. To further uncover the potential of LLMs in code translation, we propose two methods: (1) intermediary translation which selects an intermediary language between the source and target ones; and (2) self-training which fine-tunes LLMs on self-generated parallel data. Evaluated with CodeLlama-13B, our approach yields an average improvement of 11.7\% computation accuracy on Python-to-other translations. Notably, we interestingly find that Go can serve as a lingua franca for translating between any two studied languages.


\end{abstract}


\section{Introduction}
$\let\thefootnote\relax\footnotetext{
* Both authors contributed equally to this research.

$\dagger$ Beijun Shen is the corresponding author.}$

Code translation, which aims at migrating source code written in one programming language (PL) to another, has emerged as a rapidly growing technology in software engineering~\cite{g-transeval}. Software applications are often developed for multiple platforms, requiring the same product to be written in multiple languages. 
From the maintenance perspective, it is often demanded to migrate software to a modern language to meet the ever-growing requirements. Automatic code translation can significantly reduce developers' effort in software maintenance, improving code reliability, safety, and efficiency. 

Recently, large language models (LLMs) for code such as StarCoder~\cite{starcoder} have demonstrated promising performance in code translation~\cite{g-transeval, translation-bugs}. Through extensive pre-training on ultra-large code corpora, LLMs acquire the ability of zero-shot translation between arbitrary languages. These properties make LLMs a promising direction for code translation.


Despite the superb advances in many code comprehension and generation tasks, the potential of LLMs for code translation remains to be fully explored. Unlike other code tasks~\cite{codealpaca, octocoder}, LLMs demonstrate to be more struggling with code translation. This is because they have not been extensively pre-trained on parallel multilingual code, which code translation heavily depends on. 
Recent research~\cite{translation-bugs} has revealed that LLMs suffer from numerous kinds of translation errors, such as API mismatch, incorrect data type, and missing library imports. 
These errors suggest that there is still ample room for improving LLMs in the code translation task.


In this paper, we conduct a large-scale and comprehensive study to exploit the potential of LLMs in code translation. 
First, we build a new code translation benchmark \emph{PolyHumanEval} by extending HumanEval~\cite{octocoder} to 14 programming languages. 
We also design a rule-based tool to generate equivalent test programs for all PLs and ensure all our canonical solutions have passed those tests.
Based on PolyHumanEval, we perform over 110,000 translations with bleeding-edge code LLMs.
We find that LLMs exhibit asymmetrical capability in code translation with more proficiency in translating other languages to Python while struggling with Python-to-other translation. It also demonstrates asymmetrical capability in comprehending and generating code for the same language. Moreover, conventional pre-training and instruction tuning techniques have a marginal effect on LLM's capability in code translation. Our results also indicate that prompting LLMs with the function signature (obtained by simple rule-based conversion) of the target language enhances the code translation performance considerably. 

To further uncover the potential of LLMs in code translation, we propose two novel methods, namely, intermediary translation and self-training. The intermediary translation takes Go as a lingua franca between the source and target languages. The self-training fine-tunes LLMs on self-generated parallel code, which undergoes self-checking to ensure high quality. 
Evaluation results show that both methods enhance LLMs' ability in code translation considerably, yielding an average improvement of 11.7\% computation accuracy on Python-to-other translations with CodeLlama-13B.

Our contributions can be summarized as follows:

\begin{itemize}    
    \item 
    We build the PolyHumanEval benchmark by extending HumanEval to 14 popular PLs, and conduct a large-scale empirical study on the performance of state-of-the-art code LLM series on code translation tasks across multilingual PLs. 
    
    \item We propose novel methods of intermediary translation and self-training to unravel LLMs' potential in code translation. The results demonstrate significant improvements in the code translation capabilities of LLMs.
\end{itemize}

\section{Experimental Setup}


Our study aims to answer the following research questions:

\textbf{RQ1.} How effective are code LLMs in performing code translation?

\textbf{RQ2.} How to efficiently prompt LLMs to translate source code?

\textbf{RQ3.} What are effective ways to improve the performance of LLMs on code translation?

\subsection{Studied LLMs}
\label{models}

We select four representative open-source code LLMs, as summarized in Table \ref{LLM-Infomation}:

\IEEEpubidadjcol
\begin{itemize}
    \item \textbf{CodeLlama}~\cite{codellama} is a family of code LLMs further trained upon Llama2~\cite{llama2} using grouped query attention. We study its 7B, 13B and 34B variants.
    \item \textbf{StarCoder}~\cite{starcoder} is a code LLM built on GPT-2 structure with multi-query attention. The LLM has been trained on StarCoderData, a multilingual code dataset extracted from The Stack~\cite{the-stack}.
    \item \textbf{CodeGen}~\cite{codegen2} is a well-known family of code LLMs developed by Salesforce. We examine its latest version CodeGen2.5, which is built on Llama-7B and has been pre-trained on StarCoderData.
    \item \textbf{CodeGeeX2}~\cite{codegeex} is a code LLM with multi-query attention. The model was initialized from ChatGLM2~\cite{glm}. 
\end{itemize}

\begin{table*}[!t]
    \centering
    \caption{Statistics of The Studied Code LLMs.}
    \label{LLM-Infomation}
    \resizebox{\textwidth}{!}{%
    \begin{tabular}{l l l l}
    \toprule
        Model & Type$^1$ & Training Tasks & Training Data\\
        \midrule     
        CodeLlama-(7B/13B/34B) & base & Completion + Infilling + Long-Context-Tuning & Closed \\
        CodeLlama-(7B/13B/34B) -Python & tuned & w/ Python-Training & Closed \\
        CodeLlama-(7B/13B/34B) -Instruct & tuned & w/ Instruction-Tuning + Self-Instruct & Closed \\
        \midrule
        StarCoderBase & base & Completion + Infilling & StarCoderData \\
        StarCoder & tuned & w/ Python-Training & StarCoderData(Python) \\
        StarChat-$\alpha$ & tuned & w/ Instruction-Tuning & OpenAssistant\cite{openassistant}\\
        {OctoCoder} & {tuned} & w/ Python-Training + Instruction-Tuning & StarCoderData(Python) + OpenAssistant + CommitFT\cite{octocoder} \\
        StarCoderPlus & {tuned} & w/ Python-Training  + NL-Training & StarCoderData(Python) + RefinedWeb\cite{refinedweb} \\
        {StarChat-$\beta$} & {tuned} & w/ Python-Training + NL-Training + Instruction-Tuning & StarCoderData(Python) + RefinedWeb + OpenAssistant\\
        \midrule
        StarCoderBase-7B & base & Completion + Infilling & StarCoderData \\
        \midrule
        CodeGen2.5-7B-Multi & base & Completion + Infilling & StarCoderData \\
        CodeGen2.5-7B-Mono & tuned & w/ Python-Training & StarCoderData(Python) \\
        CodeGen2.5-7B-Instruct & tuned & w/ Python-Training + Instruction-Tuning & StarCoderData(Python) + Unknown$^2$ \\
       \midrule
        CodeGeeX2-6B & base & Completion + Infilling + Cross-File Completion & Closed \\
    \bottomrule
    \end{tabular}
    }
    \begin{tablenotes}
        \item 1 All ``tuned'' models are further trained on the ``base'' model in the same block. 
        \item 2 The technical report only mentioned that CodeGen2.5-7B-Instruct was fine-tuned on public instruction dataset but did not specify which one.
    \end{tablenotes}
    \vspace{-6pt}
\end{table*}

We provide the configurations for LLMs in Table \ref{tab:config}. 
All models are trained and evaluated on 2$\times$NVIDIA GeForce RTX 4090 and undergo int-8 quantization~\cite{int8}.


\begin{table}[]
\centering
\caption{Configurations and Hyperparameters for LLMs.}
\label{tab:config}
\resizebox{0.9\linewidth}{!}{
\begin{tabular}{l r l r}
\toprule
\multicolumn{2}{c}{\textbf{LLM Generation Settings}} &
\multicolumn{2}{c}{\textbf{LoRA Fine-tuning Settings}} \\ 
\cmidrule(lr){1-2}
\cmidrule(lr){3-4}
temperature & 0.01 & lora\_r & 16 \\
top\_p    & 0.9 & lora\_alpha & 32 \\
top\_k     & 50 & lora\_dropout & 0.05 \\
repetition\_penalty & 1.0 & learning\_rate & 5e-6 \\ 
max\_new\_tokens & 2048 & batch\_size & 16 \\ 
\bottomrule
\end{tabular}}
\vspace{-10pt}
\end{table}

\subsection{Studied Programming Languages}
\label{languages}

We select PLs according to their popularity and data availability from {The Stack~\cite{the-stack}}, which is a large code corpus for pertaining LLMs. The popularity of a language is gauged by the volume of its code data. We focus on the most popular programming languages in the corpus, including 14 PLs: C++, C\#, Dart, Go, Java, JavaScript, Kotlin, PHP, Python, Ruby, Rust, Scala, Swift, and TypeScript. 

\subsection{Benchmark}
\label{sec:eval_data}
Existing code translation benchmarks~\cite{CodeXGLUE, AVATAR, g-transeval} mainly focus on long-standing PLs (e.g., Python and Java), while paying little attention to the emerging ones (e.g., Rust and Kotlin). Though there have been previous benchmarks like HumanEvalPack~\cite{octocoder} that provide HumanEval solutions in different PLs, they do not ensure the equivalence of their solutions. 

To tackle these issues, we construct a new code translation benchmark \emph{PolyHumanEval}, a multilingual benchmark of 14 PLs. PolyHumanEval is extended from HumanEval~\cite{humaneval}, a widely-used benchmark for LLM evaluation. 
For each programming problem in the HumanEval dataset, we handcraft solutions in different PLs and validate their semantic consistency with both peer-reviewing and tests. 
To generate completely equivalent tests for different languages, we develop a rule-based tool \cite{ourrepo} to automatically generate test programs in all PLs based on problem metadata. 
This metadata is manually defined, including a function signature that employs standardized data types crafted uniformly across all languages, and the associated test cases specified by the inputs and their expected results.
Consequently, our benchmark encompasses a comprehensive collection of 2296 solutions across 14 languages for 164 problems, each solution accompanied by numerous test cases.

\subsection{Evaluation Metrics}

Following the approach of TransCoder\cite{transcoder}, we use computational accuracy (CA) as the evaluation metric for code translation. CA assesses whether the source and generated code produce the same output when given identical inputs. Specifically, only the first generated result is considered for evaluation (i.e., pass@1).


\section{Answer to RQ1: Performance}
In this section, we conduct a comprehensive study of over 110,000 translations to evaluate the capabilities of LLMs. Our study investigates 7 base models and 13 tuned LLMs across 26 translation tasks. It also takes CodeLlama-13B as a representative to delve into the LLMs' capacity for understanding and generating diverse languages on 182 tasks.

\begin{table*}[!t]
    \centering
    \caption{Translation Results of LLMs.}
    \label{RQ1-A}
    \resizebox{\textwidth}{!}{%
    \begin{tabular}{c c c c c c c c c c c c c c c c}
    \toprule
        Model & C++ & C\# & Dart & Go & Java & JS & Kotlin & PHP & Ruby & Rust & Scala & Swift & TS & Avg. \\
        \midrule
        \rowcolor{gray!8}
        \multicolumn{15}{c}{X$\rightarrow$Python } \\
        \midrule
        CodeLlama-7B & 70.12 & 78.66 & 75.00 & 79.27 & 79.27 & 76.22 & 78.66 & 78.05 & 75.00 & 81.10 & 72.56 & 78.05 & 76.83 & 76.83 \\
        CodeLlama-13B & 81.10 & \textbf{91.46} & \textbf{82.32} & 90.85 & \textbf{87.80} & \textbf{83.54} & \textbf{85.37} & \textbf{82.32} & \textbf{90.24} & 87.20 & \textbf{88.41} & 84.76 & \textbf{84.76} & \textbf{86.16} \\
        CodeLlama-34B & \textbf{87.20} & 86.59 & 80.49 & \textbf{91.46} & 84.76 & 82.93 & 80.49 & 80.49 & 84.15 & \textbf{87.80} & 82.93 & \textbf{85.37} & 81.71 & 84.34 \\
        StarCoderBase-15B & 67.68 & 77.44 & 71.34 & 75.00 & 78.66 & 81.71 & 77.44 & 76.22 & 75.00 & 73.78 & 77.44 & 80.49 & 79.27 & 76.27 \\
        StarCoderBase-7B & 74.39 & 75.00 & 70.73 & 68.29 & 74.39 & 76.83 & 71.34 & 78.05 & 67.07 & 64.02 & 67.07 & 70.73 & 76.83 & 71.90 \\
        CodeGen2.5-7B-Multi & 62.20 & 71.95 & 69.51 & 69.51 & 73.17 & 77.44 & 78.05 & 75.00 & 72.56 & 71.95 & 75.00 & 73.17 & 74.39 & 72.61 \\
        CodeGeeX2-6B & 73.17 & 75.00 & 76.22 & 77.44 & 71.95 & 76.83 & 79.27 & 78.05 & 70.73 & 80.49 & 73.17 & 79.88 & 81.10 & 76.41 \\
        \midrule
        \rowcolor{gray!8}
        \multicolumn{15}{c}{Python$\rightarrow$X} \\
        \midrule
         CodeLlama-7B & 62.20 & 71.95 & 53.66 & 57.32 & 68.29 & 70.12 & 75.00 & 57.93 & 64.02 & 51.83 & 68.29 & 60.37 & 67.07 & 63.70 \\
        CodeLlama-13B & 67.68 & 70.73 & 58.54 & 58.54 & 71.34 & 72.56 & 77.44 & 66.46 & 63.41 & 59.15 & 71.95 & 62.20 & 70.12 & 66.93 \\
        CodeLlama-34B & \textbf{75.00} & \textbf{84.76} & \textbf{70.12} & \textbf{67.07} & \textbf{79.88} & \textbf{76.22} & \textbf{78.05} & \textbf{70.12} & \textbf{66.46} & \textbf{66.46} & \textbf{76.22} & \textbf{67.07} & \textbf{73.78} & \textbf{73.17} \\
        StarCoderBase-15B & 55.49 & 67.68 & 53.05 & 53.05 & 60.37 & 67.68 & 66.46 & 63.41 & 57.93 & 48.17 & 64.02 & 34.15 & 68.29 & 58.44 \\
        StarCoderBase-7B & 57.32 & 73.17 & 52.44 & 47.56 & 72.56 & 68.29 & 69.51 & 59.76 & 47.56 & 45.12 & 56.71 & 28.05 & 70.12 & 57.55 \\
        CodeGen2.5-7B-Multi & 53.05 & 73.17 & 56.10 & 51.22 & 64.63 & 71.34 & 63.41 & 62.20 & 59.15 & 51.83 & 59.15 & 35.37 & 70.73 & 59.33 \\
        CodeGeeX2-6B & 51.83 & 59.76 & 48.17 & 53.66 & 61.59 & 59.15 & 51.22 & 54.27 & 52.44 & 37.20 & 51.83 & 36.59 & 58.54 & 52.02 \\
    \bottomrule
    \end{tabular}
    }
    \begin{tablenotes}
        \item * The best scores are in bold.
    \end{tablenotes}
\end{table*}

\subsection{Overall Performance}

We first explore how effectively code LLMs perform on code translation. 
To save computation resources while showcasing a wide array of languages, we select Python as the pivot language and investigate all its translations from and to other PLs, since Python is the most extensively studied PL in LLM-based code generation.
For each translation task, we construct a prompt that comprises the task intent (i.e., ``\texttt{Translate X code to Y}''), code in the source language, the function signature of the target language, and two in-context examples, as shown in Figure \ref{fig:rq1.1} (b).

The results are summarized in Table \ref{RQ1-A}. 
Overall, LLMs exhibit fairly high CA scores on translating code. For example, CodeLlama-13B achieves an average CA score of 76.55 on Python translation tasks. This demonstrates the vast potential of LLMs in code translation.

However, we notice that the translation performance of LLMs is highly biased: while LLMs exhibit high proficiency in translating from other languages to Python, they encounter challenges when translating from Python to other languages. For example, CodeLlama-13B achieves an average CA score of 86.16 when translating other languages to Python, while 
decreases to 66.93
when it comes to Python$\rightarrow$X translations. 
We will perform a more fine-grained analysis of this biased phenomenon in Section \ref{lang-diff}.  

Across the studied LLMs, the CodeLlama series exhibits the best performance. 
Despite having fewer parameters, CodeLlama-13B significantly outperforms StarCoderBase-15B on all tasks. 
CodeGeeX2-6B obtains comparable results with CodeLlama-7B on X$\rightarrow$Python translation but encounters difficulties in the opposite direction. 
Comparing models with similar sizes, the StarCoderBase-7B and CodeGen2.5-7B-Multi exhibit comparable performance. But both models fall behind CodeLlama-7B regarding overall performance across all tasks. We conjecture that such a difference is mainly caused by different training data. Particularly, both StarCoderBase-7B and CodeGen2.5-7B-Multi have been pre-trained on StarCoderData (311GB), which is much smaller than CodeLlama's dataset (859GB). In addition, the absence of Swift data in the StarCoderData results in subpar performance in Swift generation tasks.

\definecolor{my-blue}{rgb}{0.98, 0.98, 1.0} 
\begin{tcolorbox}[enhanced, width=\linewidth, boxrule=0.8pt, 
 left=2pt, right=2pt, top=2pt, bottom=2pt, colback=my-blue, drop fuzzy shadow=black,]
\textbf{Finding 1:} LLMs exhibit asymmetrical capability in code translation: it demonstrates proficiency on X$\rightarrow$Python tasks while struggling with Python$\rightarrow$X tasks.
\end{tcolorbox}
\vspace{1pt}

\subsection{Comparing across Different Programming Languages}
\label{lang-diff}

\begin{table*}[!t]
    \centering
    \caption{Translation Results by CodeLlama-13B between All Language Pairs.}
    \label{RQ1-B}
    \resizebox{\textwidth}{!}{%
    \begin{tabular}{c c c c c c c c c c c c c c c c}
    \toprule
       Source$\backslash$Target & C++ & C\# & Dart & Go & Java & JS & Kotlin & PHP & Python & Ruby & Rust & Scala & Swift & TS & Und. Avg. \\
        \midrule
        C++ & - & 80.49 & 70.12 & 66.46 & 83.54 & 66.46 & 76.22 & 78.66 & 81.10 & 67.68 & 60.37 & 68.90 & 62.80 & 79.27 & 72.46 \\
        C\# & 71.34 & - & 64.02 & 67.07 & 73.78 & 79.27 & 79.88 & 72.56 & 91.46 & 78.05 & 57.93 & 76.83 & 64.02 & 82.32 & 73.73 \\
        Dart & 71.34 & 68.90 & - & 66.46 & 64.63 & 78.66 & 72.56 & 72.56 & 82.32 & 85.37 & 53.66 & 67.68 & 59.15 & 81.10 & 71.11 \\
        Go & 79.27 & 85.37 & 70.73 & - & 92.07 & 79.27 & 81.10 & 82.93 & 90.85 & 75.61 & 72.56 & 69.51 & 67.07 & 81.71 & \textbf{79.08} \\
        Java & 69.51 & 82.93 & 61.59 & 68.29 & - & 79.88 & 71.95 & 70.12 & 87.80 & 82.93 & 58.54 & 75.61 & 61.59 & 81.71 & 73.27 \\
        JS & 75.00 & 78.66 & 60.98 & 68.90 & 66.46 & - & 72.56 & 67.07 & 83.54 & 76.22 & 53.66 & 60.98 & 56.10 & 98.78 & 70.69 \\
        Kotlin & 71.34 & 78.66 & 62.20 & 62.20 & 65.24 & 71.34 & - & 71.34 & 85.37 & 74.39 & 62.20 & 75.00 & 65.24 & 71.34 & 70.45 \\
        PHP & 71.95 & 74.39 & 65.85 & 61.59 & 75.61 & 79.88 & 79.27 & - & 82.32 & 81.10 & 56.10 & 68.29 & 62.80 & 81.10 & 72.33 \\
        Python & 67.68 & 70.73 & 58.54 & 58.54 & 71.34 & 72.56 & 77.44 & 66.46 & - & 63.41 & 59.15 & 71.95 & 62.20 & 70.12 & \underline{66.93} \\
        Ruby & 71.95 & 78.66 & 62.80 & 65.24 & 75.61 & 80.49 & 79.27 & 76.22 & 90.24 & - & 53.66 & 73.17 & 61.59 & 77.44 & 72.80 \\
        Rust & 73.17 & 81.71 & 58.54 & 69.51 & 77.44 & 76.22 & 80.49 & 75.61 & 87.20 & 70.73 & - & 72.56 & 60.98 & 75.00 & 73.78 \\
        Scala & 70.73 & 81.10 & 67.07 & 66.46 & 78.05 & 76.83 & 82.32 & 78.05 & 88.41 & 78.66 & 60.37 & - & 60.98 & 75.00 & 74.16 \\
        Swift & 69.51 & 78.05 & 59.15 & 60.37 & 71.34 & 78.66 & 74.39 & 69.51 & 84.76 & 67.07 & 62.20 & 64.63 & - & 75.61 & 70.40 \\
        TS & 66.46 & 78.66 & 59.76 & 67.07 & 68.29 & 98.78 & 76.22 & 70.12 & 84.76 & 76.83 & 55.49 & 64.63 & 58.54 & - & 71.20 \\
        \midrule
        Gen. Avg. & 71.48 & 78.33 & 63.18 & 65.24 & 74.11 & 78.33 & 77.21 & 73.17 & \textbf{86.02} & 75.23 & \underline{58.91} & 69.98 & 61.77 & 79.27 & - \\
    \bottomrule
    \end{tabular}
    }
    \begin{tablenotes}
        \item * The scores in bold denote the highest across rows or columns, while the underlined scores denote the lowest.
    \end{tablenotes}
    \vspace{-8pt}
\end{table*}

To further investigate LLMs' capability of comprehending and generating different PLs, we take CodeLlama-13B as a representative model to perform translation tasks between all language pairs. Table~\ref{RQ1-B} summarizes the results. The comprehending (generation) score of each language is denoted as the average CA score of all translation tasks where it is the source (target). 

In terms of comprehending capability, we figure out that CodeLlama-13B attains the highest score on Go, while surprisingly gaining the lowest score on Python. One possible explanation for this counter-intuitive phenomenon is that Python has distinctive linguistic features (e.g., list comprehension) that make it difficult to understand. In comparison, Go is a procedural language, emphasizing the use of basic conditional and loop statements, which makes it easier to understand.

We surprisingly note that LLMs can understand unseen languages. For example, StarCoderBase and CodeGen2.5 models are not trained with Swift code, and thus they are incapable of translating Python to Swift. However, they achieve a competitive CA score in translating Swift to Python. 
Combined with Finding \#1 (biased capability), we believe that LLM's capabilities of comprehending and generating a specific language are imbalanced. LLMs could be skilled at generating a certain language while poor at comprehending it, and vice versa.

In terms of generation capability, however, CodeLlama-13B achieves a substantially higher score on Python than other languages, in alignment with the previous finding that LLMs are proficient at generating Python code. Meanwhile, it has the lowest score on Rust. We attribute such a difference to the linguistic features. For example, Rust has a strict syntax on ownership and mutability, which is difficult for LLMs to comprehend and generate, while Python has a freer and more human-like syntax.


\definecolor{my-blue}{rgb}{0.98, 0.98, 1.0} 
\begin{tcolorbox}[enhanced, width=\linewidth, boxrule=0.8pt, 
 left=2pt, right=2pt, top=2pt, bottom=2pt, colback=my-blue, drop fuzzy shadow=black,]
\textbf{Finding 2:} LLMs exhibit asymmetrical capabilities in comprehending and generating code for the same language. Among source languages to translate, Go is the easiest while Python is the hardest to understand. Among target languages, Python is the easiest while Rust is the hardest to generate. 
\end{tcolorbox}
\vspace{1pt}

\subsection{Comparing across Base and Tuned LLMs}

\begin{table}[!t]
    \centering
    \caption{CA Differences between Base and Tuned LLMs.}
    \label{RQ1-C}
    \resizebox{0.48\textwidth}{!}{
    \begin{tabular} {l c c c c c c c }
    \toprule
    \multirow{2}{*}{Model} & \multirow{2}{*}{Tuning$^1$} & \multicolumn{2}{c}{X$\rightarrow$Python} & \multicolumn{2}{c}{Python$\rightarrow$X} & \multicolumn{2}{c}{Total}
    \\
    \cmidrule(lr){3-4}
    \cmidrule(lr){5-6}
    \cmidrule(lr){7-8}
    & & Avg. & $\Delta$ & Avg. & $\Delta$ & Avg. & $\Delta$ \\
    \midrule
    CodeLlama-7B & - & 76.83 & - & 63.70 & - & 70.26 & - \\
    CodeLlama-7B-Python & Py & 82.69 & \better{+5.86} & 58.11 & \worse{-5.58} & 70.40 & \better{+0.14} \\
    CodeLlama-7B-Instruct & In & 80.40 & \better{+3.57} & 63.46 & \worse{-0.23} & 71.93 & \better{+1.67} \\
    \midrule
    CodeLlama-13B & - & 86.16 & - & 66.93 & - & 76.55 & - \\
    CodeLlama-13B-Python & Py & 86.77 & \better{+0.61} & 62.43 & \worse{-4.50} & 74.60 & \worse{-1.95} \\
    CodeLlama-13B-Instruct & In & 85.84 & \worse{-0.33} & 66.84 & \worse{-0.09} & 76.34 & \worse{-0.21} \\
    \midrule
    CodeLlama-34B & - & 84.34 & - & 73.17 & - & 78.75 & - \\
    CodeLlama-34B-Python & Py & 82.32 & \worse{-2.02} & 69.14 & \worse{-4.03} & 75.73 & \worse{-3.02} \\
    CodeLlama-34B-Instruct & In & 83.26 & \worse{-1.08} & 68.57 & \worse{-4.60} & 75.92 & \worse{-2.84} \\
    \midrule
    StarCoderBase & - & 76.27 & - & 58.44 & - & 67.35 & - \\
    StarCoder & Py & 78.71 & \better{+2.44} & 55.16 & \worse{-3.28} & 66.93 & \worse{-0.42} \\
    StarChat-$\alpha$ & In & 73.50 &\worse{-2.77} & 59.01 & \better{+0.56} & 66.25 & \worse{-1.10} \\
    OctoCoder & Py, In & 77.49  & \better{+1.22} & 50.61 & \worse{-7.83} & 64.05 & \worse{-3.31} \\
    StarCoderPlus & Py, NL & 68.53  & \worse{-7.74} & 44.23 & \worse{-14.21} & 56.38 & \worse{-10.98} \\
    StarChat-$\beta$ & Py, NL, In & 69.18  & \worse{-7.08} & 47.80 & \worse{-10.65} & 58.49 & \worse{-8.87} \\
    \midrule
    CodeGen2.5-7B-Multi & - & 72.61 & - & 59.33 & - & 65.97 & - \\
    CodeGen2.5-7B-Mono & Py & 68.43 & \worse{-4.18} & 24.30 & \worse{-35.04} & 46.36 & \worse{-19.61} \\
    CodeGen2.5-7B-Instruct & Py, In & 75.00 & \better{+2.39} & 31.43 & \worse{-27.91} & 53.21 & \worse{-12.76} \\
    \bottomrule
    \end{tabular}}
    \begin{tablenotes}
        \item 1 `Py', `In', and `NL' are abbreviations for further Python pre-training, instruction tuning, and further NL pre-training.
    \end{tablenotes}
    \vspace{-8pt}
\end{table}

The releases of an LLM are often accompanied by their continual pre-trained or instruction-tuned variants, which often show better performance on certain tasks like code generation.
To verify whether they work in code translation, we compare the performance of all 13 tuned variants to that of their base models. The results are summarized in Table \ref{RQ1-C}. 

\subsubsection{Effect of Further Python Pre-training}
We find that further pre-training on Python is effective on smaller LLMs, such as CodeLlama-7B-Python and StarCoder, on X$\rightarrow$Python tasks. Comparatively, they have a negative effect on Python$\rightarrow$X tasks.
A possible reason is that further training in Python intensifies LLMs' knowledge of Python while weakening other languages, thus harming the polyglot capability of LLMs.

\subsubsection{Effect of Instruction Tuning}
Instruction tuning also hampers code translation. For example, instruction tuning on both CodeLlama-13B and CodeLlama-34B leads to lower scores on all tasks. This is possibly caused by the discrepancy of objectives between fine-tuning and the final translation task. Instruction tuning aims at teaching LLMs to follow natural language instructions, which are closer to NL-to-code than code-to-code tasks.

\subsubsection{Effect of Further NL Pre-training}
LLMs (e.g., StarCoderPlus, StarChat-$\beta$) further pre-trained on NL corpora demonstrate significantly lower performance compared with their base models. This is probably because LLMs gradually forget PL knowledge during the process of NL pre-training.

\definecolor{my-blue}{rgb}{0.98, 0.98, 1.0} 
\begin{tcolorbox}[enhanced, width=\linewidth, boxrule=0.8pt, 
 left=2pt, right=2pt, top=2pt, bottom=2pt, colback=my-blue, drop fuzzy shadow=black,]
\textbf{Finding 3:}
The widely-used LLM optimization techniques, such as instruction tuning, show negative impacts on code translation, presumably due to the divergence between their objectives.
\end{tcolorbox}
\vspace{1pt}

\section{Answer to RQ2: Prompting LLMs}
\label{rq2}

\begin{figure*}[htbp]
        \centering        \includegraphics[scale=0.5, trim=0 0 0 0]{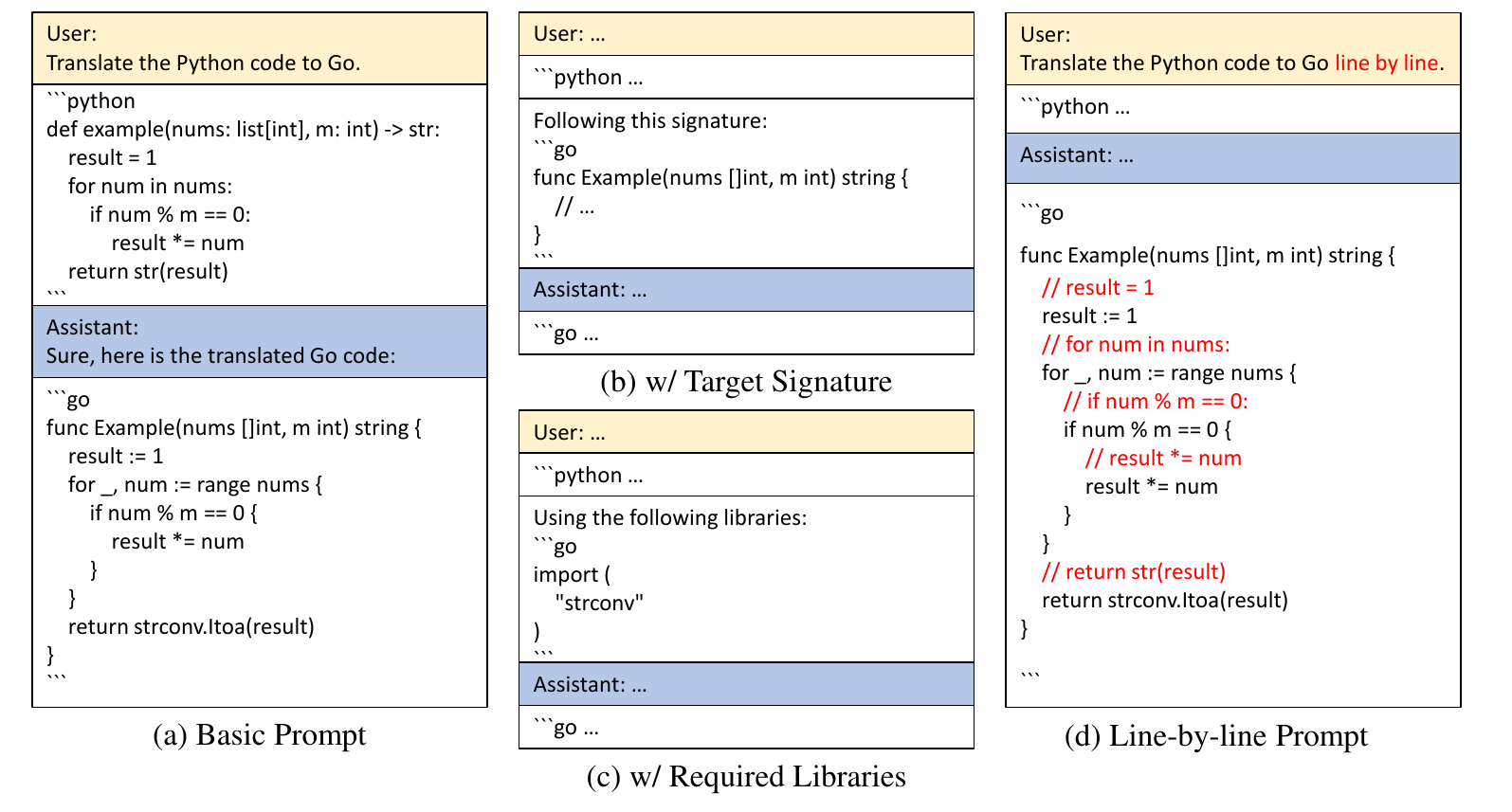}
        \caption{An Illustration of the Four Prompt Designs in Our Experiments.}
        \label{fig:rq1.1}
\end{figure*}

\begin{table*}[!t]
    \centering
    \caption{Results of Different Prompts@CodeLlama-13B.}
    \label{RQ0-A}
    \begin{tabular}{c c c c c c c c c c c c c c c c}
    \toprule
        Prompt & C++ & C\# & Dart & Go & Java & JS & Kotlin & PHP & Ruby & Rust & Scala & Swift & TS & Avg. \\
        \midrule
        \rowcolor{gray!8}
        \multicolumn{15}{c}{X$\rightarrow$Python } \\
        \midrule
        basic & 78.05 & 85.98 & 75.61 & 87.80 & 85.98 & 82.32 & 81.71 & 81.10 & 82.93 & 82.93 & 85.37 & 80.49 & 81.71 & 82.46 \\
        w/ target signature & 81.10 & \textbf{91.46} & 82.32 & \textbf{90.85} & \textbf{87.80} & \textbf{83.54} & 85.37 & \textbf{82.32} & \textbf{90.24} & 87.20 & \textbf{88.41} & \textbf{84.76} & \textbf{84.76} & \textbf{86.16}  \\
        w/ required libraries & \textbf{81.71} & 86.59 & \textbf{85.37} & 90.24 & \textbf{87.80} & 81.71 & \textbf{85.98} & 80.49 & 87.20 & \textbf{89.63} & 85.98 & 81.71 & \textbf{84.76} & 85.32 \\
        line-by-line & 77.44 & 88.41 & 69.51 & 85.37 & 82.32 & 79.88 & 79.27 & 78.05 & 81.71 & 72.56 & 80.49 & 81.10 & 77.44 & 79.50  \\
        \midrule
        \rowcolor{gray!8}
        \multicolumn{15}{c}{Python$\rightarrow$X } \\
        \midrule
        basic & 59.76 & 66.46 & 57.32 & 54.88 & 68.90 & 70.12 & 74.39 & 64.63 & 59.15 & \textbf{59.15} & 68.29 & 56.10 & 65.85 & 63.46  \\
        w/ target signature & \textbf{67.68} & 70.73 & \textbf{58.54} & 58.54 & \textbf{71.34} & \textbf{72.56} & \textbf{77.44} & \textbf{66.46} & \textbf{63.41} & \textbf{59.15} & \textbf{71.95} & \textbf{62.2} & \textbf{70.12} & \textbf{66.93} \\
        w/ required libraries & 64.63 & \textbf{77.44} & 50.61 & \textbf{59.15} & 64.63 & \textbf{72.56} & 71.34 & 65.24 & 57.32 & 56.10 & \textbf{71.95} & 54.27 & 64.02 & 63.79  \\
        line-by-line & 60.37 & 67.07 & 57.32 & 48.78 & 66.46 & 70.12 & 64.02 & 61.59 & 60.98 & 53.05 & 67.68 & 55.49 & 69.51 & 61.73 \\
    \bottomrule
    \end{tabular}
    \begin{tablenotes}
        \item * The best scores are in bold.
    \end{tablenotes}
    \vspace{-8pt}    
\end{table*}

Given that prompting is the most straightforward way of engaging with LLMs, we explore the design of prompts to harness the potential of LLMs for code translation.
We focus on two key factors in prompt design, namely, the content of the prompt and the number of demonstration examples in the prompt, to figure out the optimal setting.

As illustrated in Figure~\ref{fig:rq1.1}, we design a straight prompt~(a) that merely contains an intention: ``\texttt{translate the X code to Y}'' followed by code in the source language.
Based on the straight prompt, we augment with three variants~(b-d) using two popular prompt engineering techniques:

\begin{itemize}
\item \emph{Prompt priming} provides additional context information to LLMs, including the function signature of the target code (Figure~\ref{fig:rq1.1}.b) and required libraries (Figure~\ref{fig:rq1.1}.c), 
which have been shown to be effective in code generation tasks \cite{chen2023effectiveness}. We obtain the function signature by rule-based conversion and extracting library import statements of each question from our handcrafted solutions.

\item \emph{Chain-of-thought prompting}~\cite{chain-of-thought} asks an LLM to solve a problem via multiple intermediate steps. We prompt LLMs with examples of line-to-line alignment between the source and target languages (Figure~\ref{fig:rq1.1}.d), and ask them to translate the source code line by line.
\end{itemize}

We conduct experiments on CodeLlama-13B under a 2-shot setting. The results are presented in Table~\ref{RQ0-A}. 
We observe that providing more contextual hints to LLMs can greatly enhance the performance. Particularly, function signature in the target language plays a critical role in LLM-based code translation. The average CA score increases by 3.70 (3.47) in X$\rightarrow$Python (Python$\rightarrow$X) tasks. Prompts with required libraries demonstrate efficacy on X$\rightarrow$Python while having a marginal impact on Python$\rightarrow$X tasks. Meanwhile, line-by-line prompting leads to a decrease in the CA score.

\begin{table}[!t]
    \small
    \centering
    \caption{Results under Different N-shot Settings.}
    \label{RQ0-B}
    \begin{tabular}{c c c c c}
    \toprule
        Task & Model & 0-shot & 1-shot & 2-shot \\
        \midrule
        \multirow{3}{*}{\shortstack{X$\rightarrow$Python \\ Average}} &
        CodeLlama-7B & 76.17 & \textbf{79.04} & 76.83 \\
        & CodeLlama-13B & 82.36 & 85.60 & \textbf{86.16} \\
        & CodeLlama-34B & 67.68 & 80.86 & \textbf{84.34} \\
        \midrule
        \multirow{3}{*}{\shortstack{Python$\rightarrow$X \\ Average}} &
        CodeLlama-7B & 60.27 & 63.41 & \textbf{63.70}  \\
        & CodeLlama-13B & 63.46 & 64.54 & \textbf{66.93} \\
        & CodeLlama-34B & 50.98 & 69.56 & \textbf{73.17}  \\
        \bottomrule
    \end{tabular}

    \begin{tablenotes}
        \item * The best scores are in bold.
    \end{tablenotes}
    \vspace{-8pt}
\end{table}

To understand the impact of demonstration examples, we evaluate LLM performance under 0-shot, 1-shot, and 2-shot settings, respectively, using the best prompt template, i.e., prompt with target signature.
We adopt CodeLlama (7B, 13B, and 34B) to perform the translation tasks.
Table \ref{RQ0-B} shows that in most cases, 2-shot translation achieves the best performance. 
Especially, CodeLlama-34B gains 43.5\% (24.6\%) higher CA with 2-shot settings than 0-shot on Python$\rightarrow$X (X$\rightarrow$Python) tasks. 

\definecolor{my-blue}{rgb}{0.98, 0.98, 1.0} 
\begin{tcolorbox}[enhanced, width=\linewidth, boxrule=0.8pt, 
 left=2pt, right=2pt, top=2pt, bottom=2pt, colback=my-blue, drop fuzzy shadow=black,]
\textbf{Finding 4:} Prompt priming and examples are the effective ways to unlock LLM capacity on code translation. Among various priming techniques, providing the function signature (obtained by simple rule-based conversion) of the target language yields the optimal translation. 
\end{tcolorbox}
\vspace{1pt}

\section{Answer to RQ3: Improvement}
To further uncover the potential of LLMs, we propose two novel LLM optimization methods: intermediary translation and self-training.

\subsection{Intermediary Translation}

\begin{figure*}[htbp]
        \centering
        \includegraphics[scale=0.46, trim=-10 0 30 0]{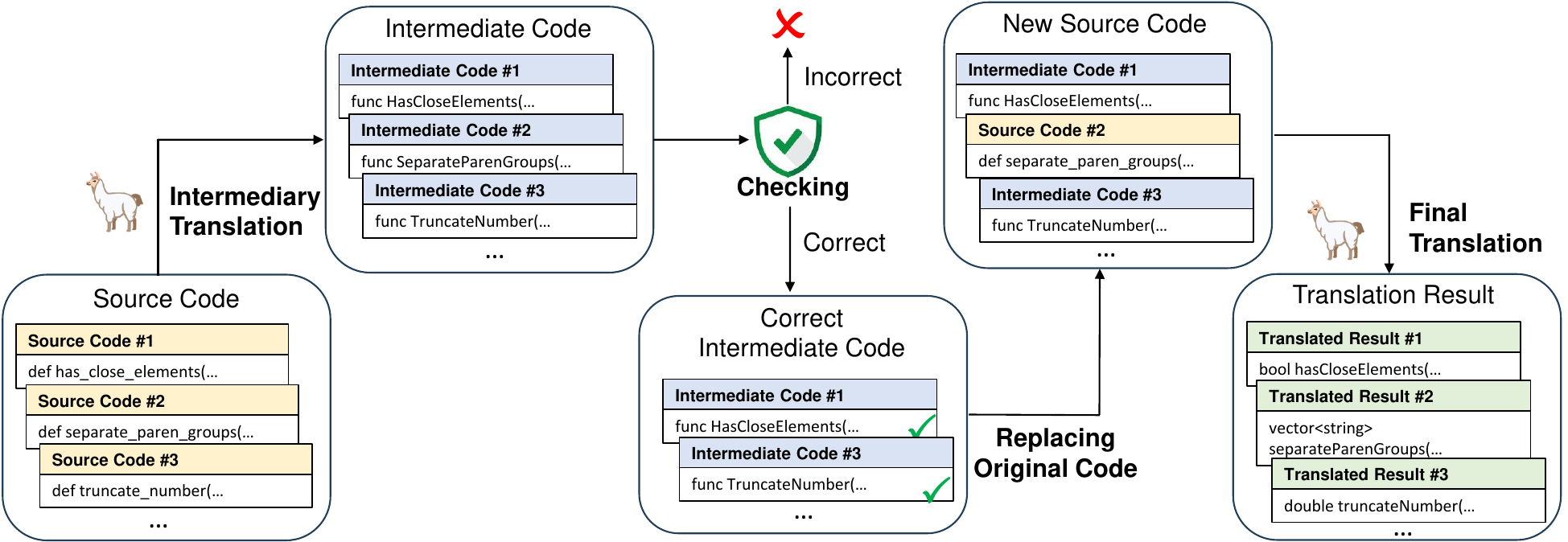}
        \caption{An Illustration of Intermediary Translation. }
        \label{fig:rq3.2}
        \vspace{-8pt}
\end{figure*}

Inspired by prior studies on multi-step LLM inferences~\cite{chain-of-thought,multi-step-reasoning}, we propose a two-step translation method, as illustrated in Figure \ref{fig:rq3.2}. For a given code snippet to be translated, we first ask the LLM to transform it into an intermediate translation result. The translation then undergoes a test using the corresponding test cases. If correct, the LLM then translates the intermediate result to the target language. 

We design two techniques for generating the intermediate translation: 

1) \textit{Style transfer}, which involves instructing the LLM to transform code written in the source language into code that adopts a comparable style to the target language. For example, when LLM performs Python$\rightarrow$Java translation, it first translates the source code into a pure procedural style, e.g., converting the list comprehension and functional programming APIs into simple `\texttt{if}' and `\texttt{for}' statements. 

2) \textit{Intermediary language}, which asks the LLM to translate the code to a third-party language that acts as a lingua franca in the programming world. The idea is inspired by natural language translation scenarios where we often use English for communication across different natural languages.


\subsection{Self-Training}

\begin{figure*}[htbp]
        \centering
        \includegraphics[scale=0.38, trim=-10 0 30 0]{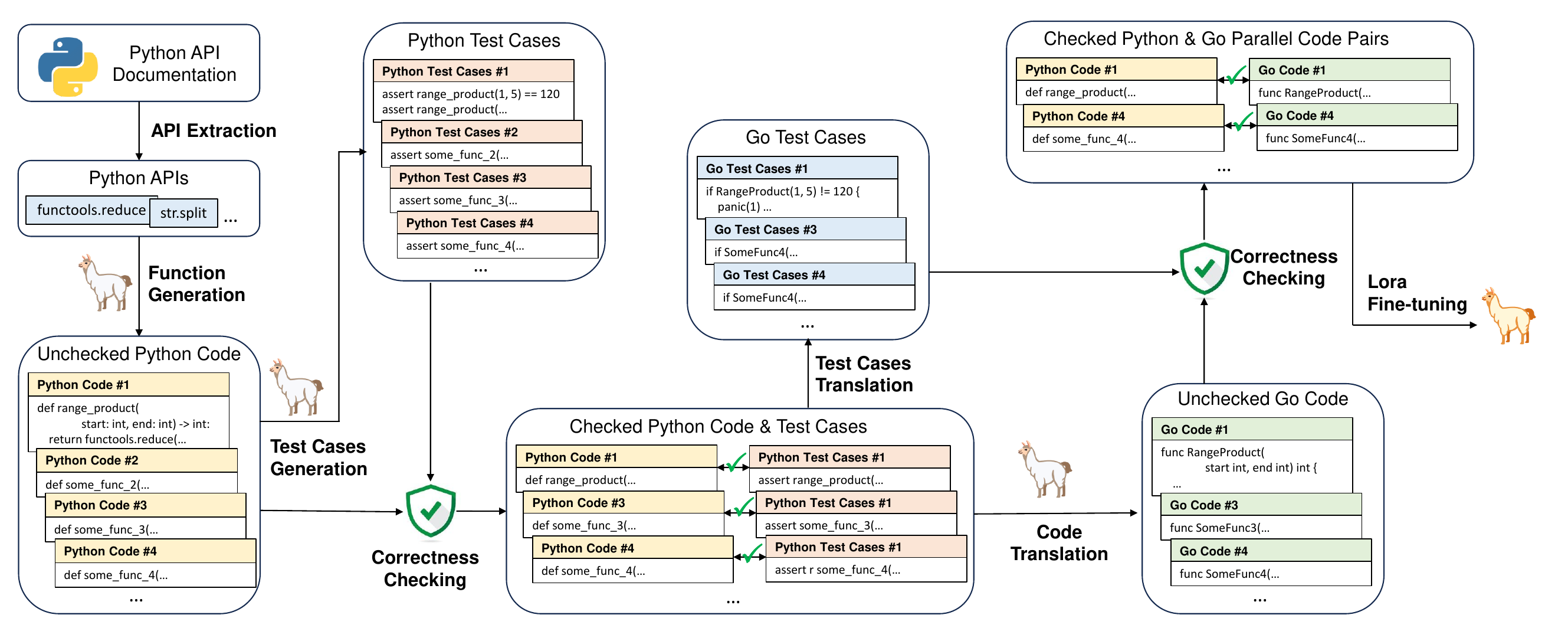}
        \caption{An Illustration of Self-training.}
        \label{fig:rq3.3}
\end{figure*}

To alleviate the data-hungry issue of parallel multilingual code, we propose a self-training method for code translation. We harness the capabilities of the LLM to generate source code, test cases, and parallel code on target languages for fine-tuning itself. 
Take Python$\rightarrow$Go translation in Figure \ref{fig:rq3.3} as an example. To enhance the variety of code data, we first extract a range of APIs from Python documentation, including built-in APIs and those from standard libraries. 
Next, for each API, we ask the LLM to generate functions that invoke this API. We also ask the LLM to construct 5 corresponding test cases for each function and filter out functions that fail to pass the test. 
The remaining functions are translated to Go by prompting the LLM. The translated Go functions are also filtered through test cases. We convert the Python test cases in the previous step to Go with our rule-based test case generation tool. 
Finally, we obtain an augmented data set with verified Python-Go code. Subsequently, LLMs are fine-tuned on the augmented dataset.

\subsection{Evaluation}

\subsubsection{Effect of Intermediary Translation}
\label{Intermediary-Translation-Result}
We conduct experiments with CodeLlama-13B on Python$\rightarrow$X, which have been shown to be the most challenging translation tasks. During the process of style transfer, we construct 2-shot prompts to instruct LLMs to convert 
Python source code into a pure procedural style. 

As shown in Table \ref{RQ3.2-A}, intermediary translation techniques, including style transfer and translation via intermediary languages, have achieved remarkable improvements overall. This can be attributed to their ability to assist LLMs to better understand the source language. C++, C\#, Go and Java are effective intermediary languages that could improve performance. Among them, using Go as the intermediary language gets the best result, which is identical to the conclusion that Go is the easiest to understand (Section~\ref{lang-diff}).

We then conduct experiments on three CodeLlama models with style transfer and intermediary translation via Go, and the results are shown in Table \ref{RQ3.2-B}.
Both methods are effective across all LLMs, improving CA by 2.58$\sim$4.73 points. While intermediary translation via Go consistently enhances the performance across all tasks, the impact of style transfer varies depending on the specific task. Languages such as Ruby and Scala do not reap the benefits of procedural style transfer, potentially because they are more aligned with the functional programming paradigm.
This underscores the profound impact of source code style on the complexity of translation.

\definecolor{my-blue}{rgb}{0.98, 0.98, 1.0} 
\begin{tcolorbox}[enhanced, width=\linewidth, boxrule=0.8pt, 
left=2pt, right=2pt, top=2pt, bottom=2pt, colback=my-blue, drop fuzzy shadow=black,]
\textbf{Finding 5:} Intermediary translation effectively enhances the code translation capabilities of LLMs, and Go can serve as a lingua franca for translating between any two studied PLs.
\end{tcolorbox}
\vspace{1pt}

\begin{table*}[!t]
    \centering
    \caption{ Results of Intermediary Translations@CodeLlama-13B on Python$\rightarrow$X Tasks.}
    \label{RQ3.2-A}
    \resizebox{\textwidth}{!}{
    \begin{tabular}{l c c c c c c c c c c c c c c}
    \toprule
        Approach & C++ & C\# & Dart & Go & Java & JS & Kotlin & PHP & Ruby & Rust & Scala & Swift & TS & Avg. \\
        \midrule
        Baseline & 67.68 & 70.73 & 58.54 & 58.54 & 71.34 & 72.56 & 77.44 & 66.46 & 63.41 & 59.15 & 71.95 & 62.20 & 70.12 & 66.93 \\ 
        w/ ST & 67.07 & 78.66 & 61.59 & 61.59 & 78.05 & 76.22 & 76.22 & \textbf{75.61} & 61.59 & 62.20 & 69.51 & 65.24 & 72.56 & \better{69.70} \\ 
        w/ IL(C++) & - & 75.61 & 63.41 & \textbf{67.07} & \textbf{78.66} & 75.61 & 72.56 & 73.78 & 67.68 & \textbf{62.80} & 67.68 & 61.59 & 75.00 & \better{69.93} \\ 
        w/ IL(C\#) & 67.07 & - & 61.59 & 63.41 & 66.46 & 75.00 & 80.49 & 71.95 & 75.00 & 60.37 & \textbf{76.22} & 65.85 & 75.61 & \better{69.98} \\ 
        w/ IL(Dart) & 63.41 & 69.51 & - & 61.59 & 64.63 & 72.56 & 70.73 & 68.90 & 75.61 & 53.66 & 60.37 & 57.32 & 67.68 & \worse{64.96} \\ 
        w/ IL(Go) & \textbf{74.39} & 76.22 & \textbf{70.12} & - & \textbf{78.66} & 72.56 & \textbf{81.10} & 71.95 & 70.12 & 59.76 & 71.95 & \textbf{67.07} & 74.39 & \better{\textbf{71.29}} \\ 
        w/ IL(Java) & 65.85 & 73.17 & 62.80 & 66.46 & - & \textbf{78.05} & 73.78 & 73.17 & \textbf{76.22} & 60.98 & 71.34 & 61.59 & \textbf{76.22} & \better{70.07} \\ 
        w/ IL(JavaScript) & 67.07 & 66.46 & 56.10 & 57.32 & 67.07 & - & 70.12 & 68.29 & 68.90 & 56.71 & 66.46 & 59.15 & 75.00 & \worse{65.48} \\ 
        w/ IL(Kotlin) & 69.51 & 68.29 & 57.32 & 59.76 & 64.02 & 75.00 & - & 68.29 & 64.02 & 55.49 & 64.63 & 55.49 & 72.56 & \worse{65.52} \\ 
        w/ IL(PHP) & 60.37 & 66.46 & 58.54 & 57.93 & 70.73 & 75.61 & 73.78 & - & \textbf{76.22} & 52.44 & 68.29 & 59.15 & 70.12 & \worse{65.85} \\ 
        w/ IL(Ruby) & 67.07 & 73.17 & 57.93 & 62.20 & 70.73 & 73.78 & 71.34 & 68.90 & - & 55.49 & 70.12 & 60.37 & 70.73 & \worse{66.56} \\ 
        w/ IL(Rust) & 73.17 & 71.95 & 58.54 & 59.76 & 72.56 & 71.34 & 71.95 & 70.73 & 67.68 & - & 68.29 & 57.32 & 70.73 & \better{67.17} \\ 
        w/ IL(Scala) & 68.29 & \textbf{79.88} & 52.44 & 58.54 & 72.56 & 75.00 & 68.29 & 69.51 & 70.73 & 57.32 & - & 59.76 & 70.73 & \better{67.31} \\ 
        w/ IL(Swift) & 67.07 & 71.34 & 57.93 & 60.98 & 71.34 & 72.56 & 75.00 & 63.41 & 60.98 & 55.49 & 64.63 & - & 67.68 & \worse{65.43} \\ 
        w/ IL(TypeScript) & 66.46 & 64.02 & 53.05 & 57.32 & 65.85 & 75.61 & 71.34 & 65.85 & 67.07 & 55.49 & 67.68 & 56.10 & - & \worse{64.30} \\ 
        
    \bottomrule
    \end{tabular}
    }
    \begin{tablenotes}
        \item * The best scores are in bold. ST stands for style transfer, and IL(X) stands for translation with intermediary language X.
    \end{tablenotes}
    \vspace{-8pt}
\end{table*}

\begin{table*}[!t]
    \centering
    \caption{Improved CA Scores by Intermediary Translation on Python$\rightarrow$X Tasks.}
    \label{RQ3.2-B}
    \resizebox{\textwidth}{!}{%
    \begin{tabular}{c c c c c c c c c c c c c c c c}
    \toprule
        Model & Appoach & C++ & C\# & Dart & Go & Java & JS & Kotlin & PHP & Ruby & Rust & Scala & Swift & TS & Avg. \\
        \midrule
        \multirow{2}{*}{CodeLlama-7B} & w/ ST & 0.00 & +3.05 & +7.93 & +3.66 & +0.61 & +0.61 & +3.05 & +9.75 & -4.26 & +5.49 & -0.61 & +1.22 & +3.05 & +2.58 \\ 
        & w/ IL(Go) & +1.82 & +1.83 & +9.14 & - & +2.44 & +3.05 & +3.66 & +8.53 & +2.44 & 0.00 & +1.22 & +1.22 & +3.05 & +3.20 \\ 
        \midrule
        \multirow{2}{*}{CodeLlama-13B} & w/ ST & -0.61 & +7.93 & +3.05 & +3.05 & +6.71 & +3.66 & -1.22 & +9.15 & -1.82 & +3.05 & -2.44 & +3.04 & +2.44 & +2.77 \\ 
        & w/ IL(Go) & +6.71 & +5.49 & +11.58 & - & +7.32 & 0.00 & +3.66 & +5.49 & +6.71 & +0.61 & 0.00 & +4.87 & +4.27 & +4.73 \\ 
        \midrule
        \multirow{2}{*}{CodeLlama-34B} & w/ ST & +2.44 & +1.83 & +6.71 & +4.27 & +6.71 & +4.88 & +1.83 & +10.98 & -11.58 & +4.88 & 0.00 & +3.05 & +3.05 & +3.00 \\ 
        & w/ IL(Go) & +1.22 & +1.83 & +7.93 & - & +1.22 & +2.44 & -1.83 & +8.54 & -1.22 & +4.27 & +3.66 & +3.66 & +4.27 & +3.00 \\ 
    \bottomrule
    \end{tabular}
    }
    \begin{tablenotes}
        \item * ST stands for style transfer, and IL(Go) stands for intermediary translation via Go.
    \end{tablenotes}
    \vspace{-2pt}
\end{table*}

\subsubsection{Effect of Self-Training}
We generate 20,000 Python functions among which 3,915 are verified to be correct, and then fine-tune CodeLlama-13B on three parallel code datasets:
\begin{enumerate}
    \item \textit{Verified Pass@1 Data}: We generate 1 translation for each Python function with temperature = 0.01 and obtain 2,407 verified Python-Go code pairs.
    \item \textit{Verified Pass@5 Data}: We generate 5 translations for each Python function with temperature = 0.8 and obtain 3,224 verified Python-Go code pairs. When there are multiple valid results, a random one is selected. 
    \item \textit{Unchecked Data}: We generate 1 translation for each Python function with temperature = 0.01 \emph{without} verification and obtain 3,915 Python-Go code pairs.
\end{enumerate}

Table \ref{RQ3.3-A} summarizes the evaluation results. 
For the Python$\rightarrow$Go task, fine-tuning on the self-generated data gains remarkable improvement: the verified Pass@5 (Pass@1) data achieves 14.57\% (6.25\%) more CA score. 
It shows the effectiveness of self-training with validated data. 
Such enhancement can also be transferred to generating other PLs, that is, fine-tuning on Python$\rightarrow$Go advances all of Python$\rightarrow$X tasks.

For the Go$\rightarrow$Python task, however, fine-tuning leads to worse performance. It indicates that fine-tuning on A$\rightarrow$B task may in turn degrade the performance of B$\rightarrow$A translation. A possible explanation is the different characteristics between the source code and target code generated by LLM: the source code aims for diversity and contains rich APIs, while the target code only aims for correctness and thus tends to be simple. Therefore, it eventually harms the comprehension of target PLs.
Surprisingly, translating other PLs to Python benefits from all fine-tuning procedures, as all the fine-tuning datasets include verified Python code.

\begin{table*}[htbp]
    \centering
    \caption{Results of Self-Training with Different Fine-tuning Data@CodeLlama-13B.}
    \label{RQ3.3-A}
    \begin{tabular}{c c c c c c c c c c c c c c c c}
    \toprule
        Fine-tuning Data & C++ & C\# & Dart & Go & Java & JS & Kotlin & PHP & Ruby & Rust & Scala & Swift & TS & Avg. \\
        \midrule
        \rowcolor{gray!8}
        \multicolumn{16}{c}{X$\rightarrow$Python } \\   
        \midrule
        None & 81.10 & \textbf{91.46} & 82.32 & \textbf{90.85} & 87.80 & 83.54 & 85.37 & 82.32 & 90.24 & 87.20 & 88.41 & 84.76 & 84.76 & 86.16 \\
        Unchecked Data & 84.76 & 87.20 & 82.32 & 89.63 & 87.20 & 87.20 & 87.20 & 87.80 & \textbf{90.85} & 90.85 & \textbf{90.85} & 86.59 & 86.59 & 87.62 \\
        Checked Pass@1 Data & 87.20 & 89.02 & \textbf{84.15} & 87.20 & \textbf{88.41} & \textbf{89.02} & 86.59 & 87.80 & \textbf{90.85} & \textbf{91.46} & 90.24 & 85.98 & \textbf{88.41} & \textbf{88.18} \\
        Checked Pass@5 Data & \textbf{89.02} & 89.02 & \textbf{84.15} & 87.80 & 87.80 & 87.20 & \textbf{89.63} & \textbf{89.02} & 90.24 & 89.63 & 89.63 & \textbf{87.20} & 85.98 & \textbf{88.18} \\
        \midrule
        \rowcolor{gray!8}
        \multicolumn{16}{c}{Python$\rightarrow$X } \\
        \midrule
        None & 67.68 & 70.73 & 58.54 & 58.54 & 71.34 & 72.56 & 77.34 & 66.46 & 63.41 & 59.15 & 71.95 & 62.20 & 70.12 & 66.93 \\
        Unchecked Data & 66.46 & 67.68 & 57.93 & 58.54 & 69.51 & 74.39 & 78.05 & 65.85 & 68.29 & 60.37 & 70.12 & 60.37 & 71.95 & 66.89 \\
        Checked Pass@1 Data & \textbf{68.90} & 71.34 & 63.41 & 62.20 & 71.95 & 76.22 & 81.10 & 68.29 & 71.34 & \textbf{63.41} & 71.95 & 64.63 & 75.61 & 70.03 \\
        Checked Pass@5 Data & \textbf{68.90} & \textbf{71.95} & \textbf{65.85} & \textbf{67.07} & \textbf{76.22} & \textbf{77.44} & \textbf{82.32} & \textbf{69.51} & \textbf{76.22} & 57.93 & \textbf{72.56} & \textbf{65.85} & \textbf{76.83} & \textbf{71.43} \\
    \bottomrule
    \end{tabular}
    \begin{tablenotes}
        \item * The best scores are in bold.
    \end{tablenotes}
    \vspace{-2pt}    
\end{table*}

\definecolor{my-blue}{rgb}{0.98, 0.98, 1.0} 
\begin{tcolorbox}[enhanced, width=\linewidth, boxrule=0.8pt, 
left=2pt, right=2pt, top=2pt, bottom=2pt, colback=my-blue, drop fuzzy shadow=black,]
\textbf{Finding 6:} 
Self-training significantly optimizes LLM's performance on the targeted translation task, and also boosts other translation tasks due to LLM's transfer learning capability. 
\end{tcolorbox}
\vspace{1pt}

\subsubsection{Effect of Combined Approach}
Finally, we assess the effectiveness when adopting both optimization methods. As shown in Table \ref{RQ3.3-B}, combining self-training and intermediary translation has the best performance of 74.77 average CA on Python$\rightarrow$X tasks. It gains 4.9\% higher CA than using intermediary translation only, 4.7\% more than using self-training only, and 11.7\% more than the original CodeLlama-13B.

\begin{table*}[htbp]
    \centering
    \caption{Results of Different Optimization Approaches@CodeLlama-13B on Python$\rightarrow$X Tasks.}
    \label{RQ3.3-B}
    \resizebox{\textwidth}{!}{%
    \begin{tabular}{l c c c c c c c c c c c c c c c}
    \toprule
        Approach$^1$ & C++ & C\# & Dart & Go & Java & JS & Kotlin & PHP & Ruby & Rust & Scala & Swift & TS & Avg. \\
        \midrule       
        CodeLlama-13B & 67.68 & 70.73 & 58.54 & 58.54 & 71.34 & 72.56 & 77.44 & 66.46 & 63.41 & 59.15 & 71.95 & 62.20 & 70.12 & 66.93 \\
        w/ ST & 67.07 & 78.66 & 61.59 & 61.59 & 78.05 & 76.22 & 76.22 & 75.61 & 61.59 & 62.20 & 69.51 & 65.24 & 72.56 & 69.70 \\
        w/ IL(Go) & \textbf{74.39} & 76.22 & 70.12 & 58.54$^2$ & 78.66 & 72.56 & 81.10 & 71.95 & 70.12 & 59.76 & 71.95 & 67.07 & 74.39 & 71.29 \\ 
        w/ Self-Training & 68.90 & 71.95 & 65.85 & 67.07 & 76.22 & 77.44 & 82.32 & 69.51 & 76.22 & 57.93 & 72.56 & 65.85 & 76.83 & 71.43 \\
        w/ Self-Training + ST & 70.73 & \textbf{81.1} & 64.63 & \textbf{68.90} & \textbf{80.49} & \textbf{81.71} & \textbf{84.15} & \textbf{76.83} & 68.90 & \textbf{67.68} & 69.51 & \textbf{71.34} & 80.49 & 74.34 \\
        w/ Self-Training + IL(Go) & 69.51 & 77.44 & \textbf{75.00} & 67.07$^1$ & 79.27 & 81.10 & 82.93 & 73.17 & \textbf{76.83} & 66.46 & \textbf{73.78} & 68.29 & \textbf{81.10} & \textbf{74.77} \\
    \bottomrule
    \end{tabular}
    }
    \begin{tablenotes}
        \item 1 The best scores are in bold. ST stands for style transfer, and IL(Go) stands for intermediary translation via Go. 
        \item 2 Since IL(Go) can not be applied to Python$\rightarrow$Go translation, the original scores are displayed here.
    \end{tablenotes}
    \vspace{-8pt}    
\end{table*}

\subsection{Case Study}

\begin{figure}[t]
       \centering
       \hspace*{0em}
       \includegraphics[scale=0.6, trim=0 0 0 0]{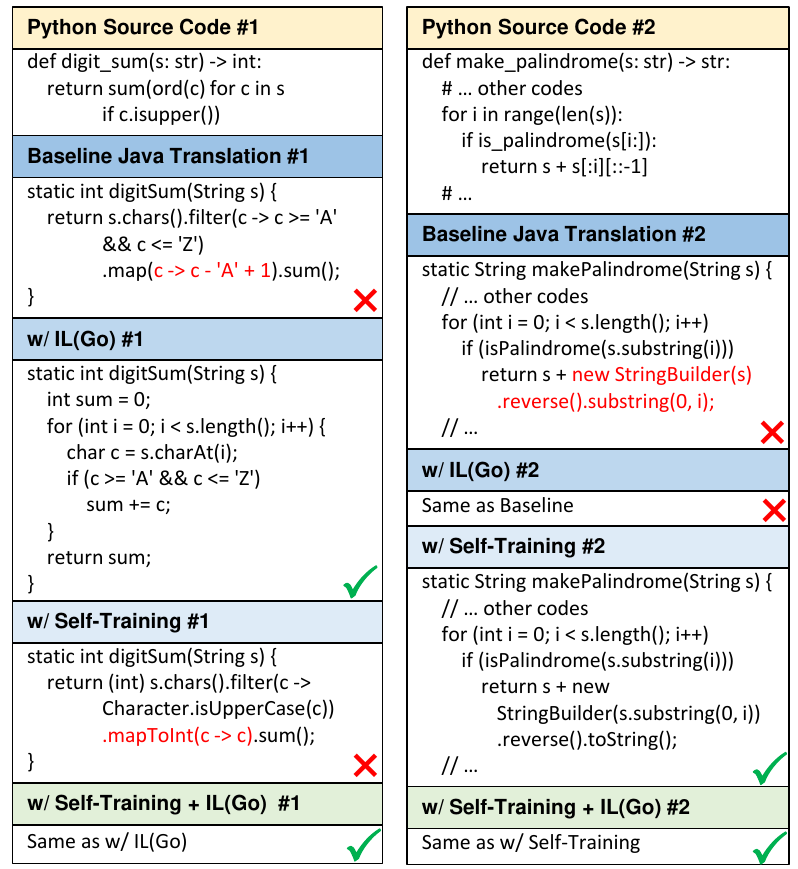}
       \caption{Examples of Python$\rightarrow$Java Translations. }
       \label{fig:case}
       \vspace{-8pt}   
\end{figure}
Figure~\ref{fig:case} shows two cases of Python$\rightarrow$Java translations by various approaches. 
In both cases, the baseline fails to adhere to the original semantics. 
Adopting intermediary translation via Go leads to an easier style of \#1 translation, but can not help example \#2.
Adopting self-training leads to correct semantics in both cases, but it fails in example \#1 due to API misuse.
Finally, the combined approach derives correct translation in both cases, achieving the best performance.

These examples demonstrate the effectiveness of our proposed methods: intermediary translation tends to generate easier code and reduces errors related to API usage; self-training makes LLMs better adhere to the original semantics; and these methods are fairly compatible with each other. More translation results are available at our online repository~\cite{ourrepo}.

\section{Discussion}

\subsection{Future Directions}
Despite the effectiveness of our optimization methods, we believe that LLM-based code translation still has a big improvement space. We delineate two future directions.

\subsubsection{Intermediary Language Selection}
Although adopting Go as lingua franca has promising results, it might not be the best intermediary language for all language pairs. 
Therefore, greater attention should be devoted to how to automatically choose the optimal intermediary language, e.g., by leveraging language embedding.

\subsubsection{High-Quality Parallel Data Generation}
Our study has demonstrated the effectiveness of the self-training method, which involves utilizing LLM itself to generate parallel code. 
However, it only ensures the correctness and API coverage of the generated code data. As for fine-tuning LLMs, the training data should also be diverse and fully reflect the usage and characteristics of programming languages. Further enhancing the quality of generated code could continually improve LLMs' performance in code translation tasks.


\subsection{Threat to Validity}
\textit{\textbf{Internal Validity.}}
The code snippets examined in our study may differ from that in real-world projects. While correctness is ensured, the handcrafted code might not reflect the programming habits of real developers, which might affect the final translation results. In the future, we will investigate the translation of real-world software projects.

\textit{\textbf{External Validity.}} 
This paper has studied four families of code LLMs, which might not represent all LLMs. Furthermore, our optimization methods are only evaluated on CodeLlama. Although these methods are applicable to any LLM, other LLMs such as GPT-4 may exhibit different performances. We leave code translation studies on other LLMs for our future work.

\section{Related Work}

Prior work has proposed rule-based, supervised, and unsupervised approaches for code translation tasks. Rule-based approaches, including CxGo \cite{cxgo} and C2Rust \cite{c2rust}, rely on conversion rules and usually produce code that lacks readability. Supervised approaches learn from existing parallel code data and produce code closer to the real-world style. It is supported by statistical translation approaches~\cite{SMTCodeTranslation2015}, tree-based neural networks~\cite{tree2tree}, and pre-trained language models~\cite{LiYGCS24}. However, their performance and generalizability are restricted by the scarcity and quality of parallel data. 

Unsupervised approaches alleviate such limitations. Lachaux et al.~\cite{transcoder} presented TransCoder, which leverages back-translation to train models without parallel data. Their later work~\cite{transcoder-st, transcoder-ir} leveraged unit tests and compiler representations for further optimization. 
However, they still require specialized training for each PL pair and thus demand substantial computation resources for generalization.

Due to the extraordinary efficacy of LLMs~\cite{gpt-4,llama2,codegen2,codegeex}, researchers have started to empirically study LLMs' code translation capability. Jiao et al.~\cite{g-transeval} proposed a 4-type taxonomy for code translation and evaluated LLMs' performance on each type of translation task. Pan et al.~\cite{translation-bugs} generalized 15 bug categories from LLMs' unsuccessful code translations. Yang et al.~\cite{Yang2024} proposed UniTrans, a code translation framework leveraging LLMs by generating test cases and repairing.
However, they studied a very limited number of PLs and only proposed prompt-crafting solutions.

Correspondingly, our work conducts a comprehensive study on code translation, covering both common and emerging PLs such as Rust, Kotlin, and Swift. We are also the first to propose systematical optimization solutions for LLMs on code translation tasks, including intermediary translation and self-training.

\section{Conclusion}
This paper conducts a large-scale empirical study to exploit the capabilities and limitations of LLMs in code translation. By investigating 4 popular LLM families on code translation tasks between 14 programming languages, we draw the conclusion that LLMs have biased and sub-optimal code translation capabilities. Further study on LLM variants proves that widely-used LLM optimization techniques are not effective in improving code translation tasks. To that end, we propose two optimization methods, namely, intermediary translation and self-training. Our evaluation on a variety of translation tasks proves their effectiveness and identifies Go as the best intermediary language. 

We have released our benchmarks, source code, experimental results, and test generation tool online \cite{ourrepo}.

\section*{Acknowledgments}
This research is supported by National Key R\&D Program of China (Grant No. 2023YFB4503802) and National Natural Science Foundation of China (Grant No. 62102244, 62232003).

\balance
\bibliographystyle{IEEEtran.bst}
\bibliography{IEEEabrv.bib,ref.bib}

\end{document}